# OPERATIONAL INTEGRATION POTENTIAL OF REGIONAL UNCREWED AIRCRAFT SYSTEMS INTO THE AIRSPACE SYSTEM

Tim Felix Sievers[1] (https://orcid.org/0000-0002-3636-424X),
Jordan Sakakeeny, Ph.D.[2] (https://orcid.org/0000-0002-8368-9127),
Nadezhda Dimitrova[2],
Husni Idris, Ph.D.[2]

[1]DLR Institute of Flight Guidance,
German Aerospace Center
Lilienthalplatz 7, 38108 Braunschweig, Germany

[2]NASA Ames Research Center
Moffett Field, CA

Contact: tim.sievers@dlr.de, jordan.a.sakakeeny@nasa.gov

**Abstract**

As part of newly developing aviation markets, fixed-wing Uncrewed Aircraft Systems (UAS) are projected to impact airspace systems and conventional air traffic in the future. The initial introduction of fixed-wing cargo UAS for regional operations is anticipated to occur at smaller under-utilized airports. Therefore, this paper assesses the integration potential of regional fixed-wing cargo UAS into the airspace system. A baseline is established to identify potential airports for cargo UAS operations in different areas. Additionally, using 2022 data, regional aircraft eligible for future cargo UAS operations are investigated. Finally, the accessibility of these regional aircraft at the identified airports was analysed. Based on the availability of current certified landing systems needed for initial UAS operations, potential airports in the areas Germany, Texas, and California for UAS operations are compared. Additionally, based on the maximum takeoff weight allowances of airport runways, current air transport operations at airports, and airspace classes, individual airports with a high potential for the introduction of initial cargo UAS operations with and without the availability of landing systems needed for UAS are identified and compared among the investigated areas. Despite a total of 173 identified airports for potential UAS operations in Germany, 376 in Texas, and 231 in California, only eleven of these airports currently have the certified landing systems needed for initial UAS operations. However, other landing system technologies that are currently under development, such as vision-based landing systems, might support UAS accessibility at the identified airports for potential UAS operations in the future.

**Keywords**

Uncrewed aircraft systems, UAS, regional air mobility, regional aircraft, air cargo, regional airport

## NOMENCLATURE

| | |
|---|---|
| AGL | Above Ground Level |
| ANSP | Air Navigation Service Provider |
| ATC | Air Traffic Control |
| ATM | Air Traffic Management |
| BTS | Bureau of Transportation Statistics |
| CAT | Category |
| CONUS | Conterminous United States |
| CTR | Controlled Traffic Region |
| C2 | Command and Control |
| EU | European Union |
| FAA | Federal Aviation Administration |
| GBAS | Ground Based Augmentation System |
| GLS | GBAS Landing System |
| GND | Ground |
| IAA | International Access Airport |
| IAP | Instrument Approach Procedures |
| IFR | Instrument Flight Rules |
| ILS | Instrument Landing System |
| MTOW | Maximum Takeoff Weight |
| NPIAS | National Plan of Integrated Airport Systems |
| PCN | Pavement Classification Number |
| P2 | Potential UAS Airport |
| P2W | Potential UAS Airport with UAS IAP |
| P2N | Potential UAS Airport without UAS IAP |
| RAM | Regional Air Mobility |
| RMZ | Radio Mandatory Zone |
| RNAV | Area Navigation |
| SES | Single European Sky |
| SVFR | Special Visual Flight Rules |
| UA | Uncrewed Aircraft |
| UAS | Uncrewed Aircraft Systems |
| UFO | Unmanned Freight Operations |
| VFR | Visual Flight Rules |

# 1. INTRODUCTION

The United States (US) and Europe both have an extensive network of airports and dense airspace. Airspace in the US is denser, on average, and airports are generally busier in terms of flight movements, enplaned passengers, and cargo per airport, than in Europe [1]. Despite the high overall number of flight movements, many US and European airports operate under capacity because travellers and air cargo are consolidated into fewer, larger aircraft on high-traffic routes via major hubs [2]. In fact, only around 0.6% of all airports in the US serve 70% of passenger flights and 1.8% of all airports in Europe are responsible for 50% of air transport services [2, 3]. Moreover, most US and European local and regional airports are increasingly under-utilized [2, 4]. The introduction of next-generation air transport systems, such as fixed-wing Uncrewed Aircraft Systems (UAS), may help to revitalize traffic at these under-utilized airports [5, 6]. UAS are highly automated aircraft without pilots on board and the most promising initial use case for the development of these increasingly autonomous aircraft systems is expected to be regional air cargo operations [6].

In recent years, congestion at major hub airports, the emergence of electric and other non-conventionally powered aircraft, and a significant pilot shortage in the regional sector have created a desire to revitalize Regional Air Mobility (RAM) and to rethink the typical hub-and-spoke air cargo model [2]. Cargo UAS provide a proving ground for increasingly autonomous technologies because they will be subject to fewer regulations in terms of safety compared to operations that transport passengers without a pilot. These fixed-wing cargo UAS will be either conversions of existing aircraft or new designs. To safely and efficiently integrate these fixed-wing UAS, whether they include new entrant aircraft or conversions, with conventional traffic, it is critical to consider and analyse the environment in which the UAS are operating. This paper aims to answer the questions, "What kind of airports are accessible to regional air cargo aircraft eligible for UAS operations, given current assumptions about technological capabilities? Where and how many of these airports are in the airspace system?" Answering these questions provides an important input to performing studies and simulations that assess the impact of cargo UAS on the airspace system and its different entities.

For the regional cargo UAS use case, it is likely that, initially, existing aircraft will be converted to UAS. Therefore, a previous study to obtain a baseline on current regional air cargo operations in the US and Europe determined three areas (Germany, Texas, and California) as good candidates for initial cargo UAS operations due to their large number of under-utilized airports and importance to the air cargo network. It was also found that turboprop aircraft dominate the regional air cargo network. In this paper, current air traffic and airport data from 2022 for Germany, Texas, and California were analysed to provide a baseline of how the introduction of fixed-wing UAS may evolve and impact airspace systems differently in different areas.

This research as shown in the following Sections 1-4 has previously been published in [7]. Section 2 reviews previous work and establishes background differences between US and European airspace. Section 3 describes the derivation of a baseline and the methodology for how that baseline will be used for comparison. Using that baseline, Section 4 compares the potential for identified airports to support UAS operations by distinguishing between different Instrument Approach Procedures (IAP) needed for initial UAS operations and Maximum Takeoff Weight (MTOW) allowances of airport runways. Section 5 assesses individual airports for potential UAS operations based on IAP availability, airspace classes, current air transport operations, and MTOW allowances. Section 6 presents concluding remarks and future work.

# 2. BACKGROUND AND PREVIOUS WORK

An airspace system can be considered as a network of different entities in controlled and uncontrolled airspace [8]. Among others, entities include airports and aviation services, procedures, and personnel managing the air traffic. When analysing and comparing US and European airspace systems, it is important to consider the different characteristics of each's Air Traffic Management (ATM) systems. The US and European ATM systems have many fundamental similarities in terms of their operational concepts. However, in Europe, 37 different national Air Navigation Service Provider (ANSP) organizations are responsible for different geographic areas, whereas in the US, airspace management is provided by one single national organization, the Federal Aviation Administration (FAA) [1, 9]. Thus, ATM in Europe occurs primarily within individual European country borders. The Single European Sky (SES) initiative was introduced by the European Union (EU) in 2004 to de-fragmentize the European airspace and jointly improve efficiencies towards safety, performance, technological contribution, human factors, and airport infrastructure [9].

## 2.1. Differences in airspaces classes

EUROCONTROL, on behalf of the EU, regularly publishes a joint report with the FAA on "ATM operational performance comparisons between the US and Europe". The latest report published in 2019 shows that, on average, the density of operations in the airspace of the Conterminous United States (CONUS) is higher than in Europe, because the US controls almost 50% more Instrument Flight Rules (IFR) flights than Europe, even though its airspace is 10% smaller geographically [1]. Table 1 provides a comparison of airspace classes in terms of being controlled by Air Traffic Control (ATC) and the separation services provided, using Germany (GER) as a European example compared to the US [10, 11, 12].

ATC is responsible for providing separation services to aircraft by ensuring minimum separation. In the US, airspace Classes A and B exist in which all flights must be separated by ATC, whereby only IFR flights are permitted in airspace Class A. In the only uncontrolled airspace, Class G, there is no separation of flights by ATC. Furthermore, there are additional rules for separation as in Special Visual Flight Rules (SVFR) operations when weather conditions are not within the Visual Flight Rules (VFR) limits [10, 12, 13].

TAB 1  Comparison of different national airspace classes

| Airspace classes[a] | Controlled | | ATC separation | |
|---|---|---|---|---|
| | GER | US | GER | US |
| A (Alpha) | -[b] | Yes | - | IFR to IFR<br>*no VFR traffic* |
| B (Bravo) | - | Yes | - | V/IFR to V/IFR |
| C (Charlie) | Yes | Yes | IFR to V/IFR | IFR to V/IFR |
| D (Delta) | Yes | Yes | IFR to IFR | IFR to IFR |
| E (Echo) | Yes | Yes | IFR to IFR | IFR to IFR |
| G (Golf) | No | No | No | No |

a. In addition to these six airspace classes, there are designated airspace areas with limitations and special use such as for military operations.

b. Unlike some other European countries, Germany has neither Class A nor B airspace in operation. France, for example, uses Class A for the airspace around its capital, Paris. Class A airspace in the United States is not around airports at all. Rather, it incorporates the airspace between 18,000 feet and 60,000 feet.

Additionally, Germany operates Radio Mandatory Zones (RMZ), which are specially created for IFR approaches at airports in uncontrolled airspace. The RMZ begins on the ground (GND) and extends to the above bordering airspace Class E, which starts between 1,000 feet and 2,500 feet Above Ground Level (AGL). Within the RMZ, carrying radio communication equipment is mandatory. However, the aircraft does not require ATC clearance for its entry, but voice communication capability and radio standby [10].

Within the different airspace classes there are further differences between Germany and the US such as the altitude AGL to which airspace extends. For example, in the US, Class D typically covers the airspace from GND to 2,500 feet AGL [11]. In Germany, Class D airspace can reach 10,000 feet Mean Sea Level and is utilized as a Controlled Traffic Region (CTR) at 32 public airports and airfields in controlled airspace [10]. In the US, however, Classes B, C, and D are utilized as controlled airspaces around airports depending on the level of flight activities (with Class B airspace being used for the busiest airports). Additionally, some towered airports in Class C or D airspace in the US become non-towered at less traffic-intensive times, such as late evening or night, and move to Class E or G airspace accordingly. For example, Waco Regional Airport (KACT) is a Class D airspace between 0600-2400 in the local time and is Class E when the tower is not operating (i.e., from 0000-0600 local time). For simplicity, airports with a physical air traffic control tower receiving separation by ATC will be counted as "towered" in this study, although some airports might not always have this tower operational.

The existence of an air traffic control tower is an important integration factor when it comes to how a remotely piloted UAS flying under IFR will integrate into the terminal airspace surrounding an airport. It is debatable whether initial entry into the airspace will occur at low-traffic towered airports or at non-towered airports. Considering towered airports first, an air traffic controller can provide separation and other services for the UAS and its remote pilot. The process of flying into and out of a towered airport will tend to be more standardized and predictable than at non-towered airports without ATC separation. However, towered airports have a tower because they are busy enough to necessitate the services an air traffic control tower provides. Integrating into a towered airport typically will mean integrating into an environment with more traffic than a non-towered airport. That additional traffic may lead to inefficient UAS operations, should the UAS not be able to integrate with the same performance as conventionally crewed aircraft. Additionally, should the UAS face an off-nominal situation, there is a much higher chance of causing disruptions with other aircraft.

Typically, non-towered airports are less busy than towered airports and therefore aircraft in their terminal area do not receive ATC separation services. Due to the "one in, one out" rule, whereby ATC will only allow one IFR aircraft operating at a non-towered airport at a time, it is guaranteed that there will be only one IFR aircraft, for example the UAS flying in or out of the airport. However, the major integration hurdle at non-towered airports is aircraft flying under VFR, especially non-cooperative VFR traffic that operates with unknown intention and thus will not actively cooperate to resolve a potential conflict. Conventionally crewed aircraft operations utilize the pilot on board to "see and avoid" other traffic. Without a pilot on board, that requirement to "see and avoid" falls to "detect and avoid" systems, which need to have minimal latency to guarantee safe operations. Because VFR aircraft may fly less predictably than IFR aircraft, a larger buffer between Uncrewed Aircraft (UA) and VFR aircraft may be needed than between UA and IFR aircraft. This increased buffer could lead to potentially inefficient integration of UAS, as they may fly a more circuitous routing to mitigate interactions with VFR. An analogy can be found in "self-driving" cars: it is relatively straightforward to automate driving on a highway, as the path is roughly fixed, and the movement of other vehicles is fairly predictable. However, "self-driving" in the city is more difficult because non-cooperatives, such as other cars pulling out of parking spots without looking, have the freedom to do what they will, making operations much more difficult to predict.

### 2.2. Differences in network and distribution of airports

Generally, it can be observed that there are a considerable number of under-utilized airports in the US and Europe, which may be candidates for initial UAS operations. In the US, about 70% of passenger flights are operated from just 30 airports (operated in the relatively busy airspace Class B), although there are over 5,000 public US airports [2]. In Europe, a similar phenomenon exists with over 2,500 less-busy airports [3, 4]. Likewise, air cargo traffic is primarily oriented around hub-and-spoke operations, namely through major international hubs [5, 14, 15]. Smaller airports are responsible for feeder traffic to the hub-and-spoke system or for point-to-point flights, with many of these less-busy airports focused on passenger transport rather than air cargo [5, 15].

Looking at the year 2022, the aforementioned trends of US airports being busier than their European counterparts, as investigated in [1], can be observed by comparing the most recent annual data from Eurostat, the statistical office of the European Union, and the US Bureau of Transportation Statistics (BTS). For commercial flight movements, multiple values, including flight movements with passengers and/or cargo on board (all operations[1]), enplaned passengers, cargo-only flight movements, and enplaned cargo in metric tonnes (t) can be found for the 34 busiest European and US airports in Table 2 [16, 17].

**TAB 2** Median values based on 34 busiest main airports by commercial flight movements in 2022

| Median value at main airports | Europe | United States |
|---|---|---|
| All operations flight movements[a] | 140,566 | 300,489 |
| Enplaned passengers | 18,752,120 | 30,750,214 |
| Cargo-only[b] flight movements | 4,433 | 9,906 |
| Enplaned cargo on board cargo-only flights (t)[b,] | 141,206 | 198,554 |

a. Flight movements refer to the sum of an arrival and departure for all national and international commercial flights that are both scheduled and non-scheduled.

b. Cargo consists of both freight and mail. "Cargo-only" flights have no passengers on board of the aircraft.

Although Table 2 indicates that the main airports in the US are busier on average, [1] states that Europe's airports have a higher number of IFR flights per active runway and airports operate closer to their capacity limits than in the US. In 2022, 8,302,587 IFR flights were operated in Europe (based on the 27 states of the European Union plus Norway and Switzerland) with 35.8% of IFR flights (2,971,433) in France and 32.7% of IFR flights in Germany (2,712,552) [18]. In the US, 15,416,640 IFR flights were handled by the FAA in FY2022[2] [19]. 13.7% of the IFR operations in the US took place at just three airports: Atlanta (KATL), Chicago O'Hare (KORD), and Dallas-Fort Worth (KDFW).

Previous analysis showed that the aircraft flying into the airports likely to be used for the introduction of cargo UAS are small, fixed-wing aircraft, also known as regional aircraft [20]. The term regional aircraft[3], in this work, refers to fixed-wing aircraft that have a payload <9 tonnes and a MTOW <25 tonnes, regardless of propulsion type. The analysis of the potential for regional air cargo operations with UAS also showed that most of the domestic[4] cargo flight movements by regional aircraft were operated within a flight distance under 1,000 kilometers [20]. 94% of the domestic cargo-only flight movements by all aircraft in Europe and 97% of the domestic cargo-only flight movements by regional aircraft in the US were operated within this flight distance. Likewise, this definition of a regional flight distance is in accordance with NASA's definition of RAM, in which regional flights are conducted in ranges between 50 and 500 nautical miles (93-926 kilometers) [21].

The same analysis proved that a higher number of flight movements by smaller regional aircraft in the US (e.g., Cessna 208 Caravan) are used to transport an equal amount of cargo (3.7 versus 3.9 million tonnes) relative to Europe, where a lower overall number of larger turboprop aircraft dominated the regional air cargo domain [20]. Considering regional turboprop aircraft types, larger aircraft are used in Europe, such as the ATR 42, ATR 72, and Embraer EMB 120. Almost 60% of European cargo flight movements were operated over longer regional flight distances between 300 and 700 kilometers. However, in the US, over 60% of cargo flight movements by regional aircraft were operated on flights less than 300 kilometers in flight distance.

Despite its high number of small commercial airports and the highest number of intra- and extra-European cargo flight movements compared to those in any other European country, Germany had fewer than 400 domestic cargo flight movements by regional aircraft in 2021 [20]. Because of the widespread existence of small commercial airports as necessary infrastructure requirements for future UAS operations in the RAM realm [2, 22], Germany can be considered a potential country for the introduction of regional cargo UAS. However, since almost no domestic cargo flights are currently operated in Germany, existing cargo flights can rarely be replaced by UAS at present. Given the benefits of highly automated cargo UAS operations such as increased flexibility in operations and reduced personnel requirements as well as lower costs [23], it can be assumed that regional cargo UAS in Germany might be introduced via additional regional cargo operations on new flight routes.

The same analysis has shown that California and Texas appear to be well suited for regional fixed-wing cargo operations in the US [20]. California, a large, populous state in the western US of similar size to Germany, and Texas, another large, populous state, in the south-central region of the US, have a similar percentage (~15%) of intra-state cargo flight movements being performed by regional aircraft (i.e., eligible for potential UAS replacement). Both Texas and California also have important large cargo sorting hubs. However, the share of airports by sizes relevant for cargo UAS operations is different in the two US states. California has a high share of small[5] airports (73, more than any other US state, except for Alaska[6]) whereas Texas has the highest share of medium-sized airports (that Eurostat refers to as other airports) compared to any other US state. These other airports, being busier than small airports, may present more challenges with respect to the integration of cargo UAS. In this context, according to Eurostat, Germany has 141 small public, commercial airports with the majority being under-utilized [16]. Germany, Texas, and California are relatively busy in terms of total number of cargo flight movements compared to other US states and European countries (see Table 3).

---

[1] The air cargo on board of "all operations flight movements" is any of cargo-only (no passengers transported), belly freight (cargo transported in the lower deck of the passenger aircraft), or combi freight (split of the main cabin of the aircraft to separate passenger seats and cargo area).

[2] FY2022, or Fiscal Year 2022, was Oct. 1, 2021, to Sept. 31, 2022.

[3] Note that, in [20], regional aircraft referred to only piston and turboprop aircraft. The term has been expanded to include jet aircraft in this work because there is a strong desire by industry to expand beyond just turboprop aircraft into larger jet aircraft.

[4] Domestic refers to flight movements within the US or within a European country.

[5] According to Eurostat, small airports are defined as airports with <15,000 annual passenger units (where one passenger unit corresponds to either one passenger or 100 kilograms of cargo); other airports have <150,000 to ≥15,000 annual passenger units, and main airports >150,000.

[6] While Alaska is a potentially very interesting use case for cargo UAS, the choice was made to study in-depth only states in the CONUS, as those results would likely be more applicable to other US states.

TAB 3  Air cargo flight movements in 2021 [20]

| Air cargo flight movements | Germany | Texas | California |
|---|---|---|---|
| Total[a] by all aircraft | 157,764 | 98,007 | 178,792 |
| Intra-state[b] by all aircraft | 15,816 | 44,504 | 138,180 |
| Total[a] by regional aircraft | 9,870 | 18,575 | 28,370 |
| Intra-state[b] by regional aircraft | 392 | 15,026 | 27,952 |

a. Refers to flight movements within the US and to intra- and extra-European cargo flight movements.
b. Intra-state refers to flight movements within a US state and within Germany.

Likewise, the investigated areas have a significant share of less-busy airports relevant for the introduction of initial UAS operations that Eurostat refers to as small and other airports. However, Germany has a comparatively low share of domestic cargo flights by regional aircraft that have the potential to become UAS by replacing current flight routes. California and Texas, on the other hand, might be prime locations with the required airport infrastructure as well as current air cargo routes for the replacement by UAS [20].

## 3. METHODOLOGY OF THE ANALYSIS OF AIRSPACE SYSTEM CHARACTERISTICS

The methodology section describes the baseline that is applied to identify potential airports for UAS operations in different areas. The current certified landing systems needed for initial UAS operations at the potential airports are introduced before concluding with the data sources used for this study.

### 3.1. Derivation of a baseline for analysis

To assess how the introduction of UAS may evolve and impact airspace systems in different areas, a baseline of accessible airports for potential UAS operations needs to be identified. In the first step, potential airports are defined based on the air transport services they provide. In the second step, potential airports are classified based on their annual number of IFR flight movements to identify less busy airports. Finally, a maximum on the number of flight movements at an airport is applied to provide a baseline of potential airports for the introduction of UAS in different areas. This methodology was applied to airports in Germany, Texas, and California.

In Germany, airports and airfields are collectively referred to as aerodromes by the German ANSP, Deutsche Flugsicherung (DFS). Here, DFS distinguishes between airports, which "require protection by a construction protection area in accordance with § 12 of the Air Traffic Act", and airfields, which do not. The construction protection area ensures that the construction of buildings within a 1.5-kilometer radius around the airport reference point, as well as on the takeoff and landing areas and safety areas, require approval by the aviation authority [24]. In this paper, for simplicity and to better align with FAA terminology, both airfields and airports will be referred to as airports.

It can be assumed that the introduction of cargo UAS will initially occur at publicly accessible airports with less busy air transport services [22]. Public airports are open for public access and do not require individual operating permissions from the airport operator as private airports do, which likely increases the flexibility of air transport operations by cargo UAS. Due to this factor and the added difficulty of interacting with military aircraft, private[7] airports, as well as military and military-public joint-use airports are excluded from consideration. Therefore, only public airports will be analysed. Public airports can be further distinguished by whether they provide commercial and/or non-commercial air transport services. Eurostat defines commercial air transport operators and commercial purposes as "scheduled or non-scheduled air transport services, or both, which are available to the public for carriage of passengers, mail, and/ or cargo" [25]. The FAA defines airports with "commercial services" as airports that are publicly owned "with at least 2,500 annual enplanements and scheduled air carrier service" [26]. In this study, the term public airport will refer to airports that are publicly accessible (regarding potential UAS operations), regardless of whether the airport currently has commercial air transport operations. For example, Heringsdorf (EDAH), despite its relatively few (688) IFR flight movements in 2022 is a public airport because it is publicly accessible for use by both commercial and general aviation aircraft [10].

According to DFS, Germany operates 15 towered International Airports of which four serve as so-called Hub airports, six as International Access Airports 1 (IAA1) and five as International Access Airports 2 (IAA2). In addition to the 15 towered International Airports, DFS defines 20 more towered airports as Regional Airports [27]. In 2022, the four German Hub airports, including Berlin (EDDB), Frankfurt (EDDF), Dusseldorf (EDDL), and Munich (EDDM,) had a median of 222,483 IFR flight movements followed by the IAA1 with a median of 77,145 annual IFR flight movements. In total, the Hub Airports and the IAA1 accounted for 87.7% of all annual IFR flight movements of all the towered airports in Germany. Looking at the IFR flight movements at IAA1 airports, Cologne/Bonn (EDDK) was the busiest IAA1 airport (119,117) and Nuremberg (EDDN) the least busy (35,714). The IAA2 had a median of 11,909 annual IFR flight movements with the greatest number of annual IFR flight movements operated at Bremen (EDDW) with 19,423 IFR flight movements and Erfurt (EDDG) as the least busy with 2,865 annual IFR flight movements. The subsequent category of airports by DFS are so-called Regional Airports with a median of 6,483 annual IFR flight movements in 2022. The most IFR flights operated at a Regional Airport was at Dortmund (EDLW) with 21,476 annual IFR flight movements, the fewest IFR flight movements operated at a Regional Airport was at Schwerin-Parchim (EDOP), with just one single annual IFR flight movement.

For the US, the FAA distinguishes between primary airports classified as Hub (large, medium, and small) and Non-hub airports, as well as between non-primary airports classified as National, Regional, Local, Basic, and Unclassified (limited activity) airports [26]. Primary airports are airports with commercial services that handle more than 10,000 passenger boardings annually. The categorization of US airports also includes special facilities such as seaplane bases or heliports, though those are excluded from this analysis. Additionally, as in Germany, the US operates military-civil joint-use airports, which, as discussed previously, will be excluded.

---

[7] German airports are distinguished by their type of operating obligation. German airports with no operating obligation (because they are privately owned) are called special airports and special airfields. Only the operator and, upon request, third parties are allowed to operate on them.

In this study, the term potential UAS airports, or P2 airports for short, is used to establish a listing of airports to which cargo UAS might fly. P2 airports include and refer to: 1) Public towered airports with annual IFR flight movements percentages under 2.2% for the given area (country/state) and 2) Public non-towered airports.

The <2.2% threshold was selected because the least busy IAA1 airport (EDDN) had 2.2% of the total annual IFR flights in Germany. Using this cut off includes the five towered IAA2 (all public) and the 20 towered Regional Airports (17 public), as defined by DFS. The towered airports that receive <2.2% of the annual IFR traffic were selected because it is unlikely that initial UAS operations will occur at the busier airports (>2.2% of IFR flight movements). Rather, it is more likely that initial UAS operations will take place at less busy airports. Additionally, there are numerous airports in Germany that are non-towered and for which there is no record of IFR and VFR flight data provided by DFS. It can be assumed that these non-towered airports have fewer flight movements than the towered airports and thus are also included in the definition of P2 airports in this study. Following these assumptions, there are 173 P2 airports (22 towered) out of 183 public airports (32 towered) in Germany.

In Texas, there are a total of 2,080 airports (383 of which are public use) with 210 commercial airports included in the National Plan of Integrated Airport Systems (NPIAS) (47 being towered). California has a total of 899 airports (242 available for public access), with 188 commercial airports included in the NPIAS (55 being towered). Applying the <2.2% cut off for towered US airports, Texas has 376 P2 airports (40 being towered) and California has 231 P2 airports (44 being towered) [28]. Similar to Germany, a significant share of current IFR flight movements is operated at the airports with annual IFR flight movements percentages >2.2% in Texas (72.4%) and in California (78.5%) [29]. For the year 2022, Fort Worth Alliance (KAFW) was the busiest P2 airport in Texas with 48,119 annual IFR flight movements and Palm Springs International (KPSP) was the busiest P2 airport in California with 47,982 annual IFR flight movements [29].

### 3.2. Introduction of current certified landing systems for initial UAS operations

IAP are used to land in Instrument Meteorological Conditions, in which visual landing is not possible. It is anticipated that UAS will utilize IAP to land at airports. However, no regulations yet exist that specify required IAP for UAS. Regulations and standards regarding UAS automatic landing capabilities and technologies will need to be put forth before UAS can fly routine operations. Nonetheless, when integrating UAS into the airspace system, it is important to consider other air traffic participants in the airspace as well as the availability of enabling procedures and technologies for initial UAS operations, such as needed IAP present at airports.

RTCA, Inc. highlights the need for automatic landing systems for UAS in its Guidance Material and Considerations for Unmanned Aircraft Systems (RTCA DO-304A, Section 2.4.6) [30]. Although automatic landing systems not based on ground based navigational aids would provide the most operational freedom for UAS, Instrument Landing System (ILS) Category (CAT) III are the only current systems that enable automatic landing[8] in nominal operations. Although no US operator has received approval for ILS CAT IIIc, with a decision height of 0 feet and a runway visual range of 0 feet, it is nonetheless the only regulatory path to automatic land at present [31]. Therefore, until such time as alternative systems are developed and certified, it is assumed that for future UAS operations at airports, the most likely current IAP for UAS is ILS CAT III, even if the existing regulations need to be adapted for UAS. Other landing systems, such as vision-based landing systems [32], are also in development, and existing Global Positioning Systems landing systems are in use in limited situations, but do not currently meet civilian aviation safety standards. Therefore, only currently certified systems are considered in this work [33]. ILS CAT III are the most stringent IAP that exist today and require the highest level of technology of all the IAP. For ILS CAT III approaches, automatic landing systems and rollout control systems are needed to control the approaching aircraft. For more information about ILS categories, see [33].

However, ILS, especially CAT III systems, do have their downsides. They are expensive to implement and maintain and they only serve a single runway end. As such, they are not installed at many airports (only 68 throughout the US [30]). Far more common are the less stringent CAT I (decision height >200 feet) and CAT II (decision height 100-200 feet) ILS. Another class of systems already in use that can be considered for future airport accessibility of UAS are Ground Based Augmentation Landing System (GBAS) Landing Systems (GLS) [34]. GLS generally need only one installation per airport. Once installed, the Global Navigation Satellite System localizer works for all runways, making it a cheaper system to install, maintain, and upgrade than ILS [35]. Of course, aircraft must be equipped with the necessary on-board systems to utilize GLS (the same is true for ILS). The categories (CAT I, II, and III) of GLS are the same as for ILS, though only CAT I and II are operational as of this writing.

Of the five different landing systems, ILS CAT I, II, and III and GLS CAT I and II, the latter three are considered UAS IAP insofar as they provide a higher potential for utilization by UAS operations. ILS CAT III is included because it is the highest-level IAP currently in use. The GLS approaches are included because they can be upgraded to CAT III more easily than ILS, once CAT III systems become available [36]. According to a SESAR estimate, full GLS rollout at airports across Europe may be achieved as early as 2036 [37]. Based on the availability of UAS IAP, this study further distinguishes between 1) P2 airports providing UAS IAP (P2W airports) and 2) P2 airports without UAS IAP (P2N airports). Thus, the airport types in this paper are:

1. **P2 Airports:** Potential UAS airports (those airports that are public use and have <2.2% of the area's IFR flight movements)

2. **P2W Airports:** P2 airports with UAS IAP (i.e., ILS CAT III or GLS CAT I/II)

3. **P2N Airports:** P2 airports without UAS IAP

---

[8] To operate in true zero visibility conditions, surface operations, such as taxiing, also need to be automated.

P2W airports have a higher potential to be initially utilized for UAS operations than P2N airports. Here, P2N airports refer to all other airports that do not currently have ILS CAT III or GLS in place, regardless of whether they provide any ILS. However, P2N airports will still be considered for future UAS operations, as they could be retrofitted with required UAS IAP at any time. Additionally, there will likely be further technological advancements that could enable UAS accessibility at these P2N airports.

### 3.3. Data sources

The data on operational airports in Germany were accessed from the Aeronautical Information Publication Germany from DFS, which are publicly accessible since January 2023 [10]. In addition to general national regulations and requirements, specific information on airports and air navigation services can be retrieved. For this paper, information was collected about the name and operational type of airport, availability of IAP, aircraft permitted by MTOW at the airports, and hours of operation for all available German operating airports. Additional data on individual German airports were accessed from DESTATIS, the German Federal Statistical Office [38, 39].

For the US, airport and runway data (e.g., landing systems available, runway weight restrictions) were gathered from the FAA's National Airspace System Resource [40]. Airport classification information was obtained from the FAA's NPIAS [41]. IFR movement counts at towered airports were sourced from the FAA's Operations Network database [29].

The statistics on commercial flight movements by regional aircraft for Europe and Germany were retrieved from Eurostat [16]. Here, a commercial flight movement represents the sum of the arrival and departure of an aircraft at an airport. In this context, specific data of the year 2022 on all domestic (i.e., flight movements within Germany) and international (i.e., flight movements between Germany and another country) flight movements for passenger and cargo air transports were analysed. The data for domestic European flight movements include data for 35 European countries, although complete data were not available for every country. Note that domestic operations within a European country can also be referred to as "intra-state" flight movements. Such intra-state flight movements for the US indicate a flight within a single US state, whereas domestic US flight movements could move between any US state or territory.

Statistics for flight movements in the United States[9] and individual airports in Texas and California were sourced from the BTS T-100 Segment data [17]. BTS data combine segment data by aircraft type, origin, destination, and airline. The data denote the number of passengers, the amount of freight, and the amount of mail per segment. Flight movements with both origin and destination outside the US are excluded from the BTS data. Generally, the flight movement values at airports calculated from the BTS data will be lower than those shown in the FAA Operational Network because only airlines with annual operating revenues of 20 million USD or more are included in the BTS data, so some smaller airlines are excluded from the database and thus this study.

### 4. ANALYSIS OF UAS ACCESSIBILITY POTENTIAL

This section focuses on the airspace system accessibility of flights eligible for UAS operations based on availability of UAS IAP. The potential to use UAS for regional aircraft at the identified P2 airports is discussed.

### 4.1. Availability of IAP at airports

Table 4 shows the count of all public and non-public airports (excluding military use airports) and P2 airports, sorted by towered and non-towered, in Germany, Texas, and California that are equipped with different categories of ILS/GLS procedures. Airports that provide multiple ILS/GLS procedures are counted in all applicable categories.

**TAB 4** Availability of ILS/GLS procedures at airports

| ILS/GLS availability | Count of airports (towered / non-towered) | | |
|---|---|---|---|
| | Germany | Texas | California |
| **Total at all airports** | **35 / 6** | **38 / 5** | **37 / 8** |
| ILS CAT I | 27 / 6 | 38 / 5 | 37 / 8 |
| ILS CAT II | 3 / 0 | 7 / 0 | 9 / 0 |
| ILS CAT III *(UAS IAP)* | 20 / 0 | 5 / 0 | 6 / 0 |
| GLS CAT I *(UAS IAP)* | 2 / 0 | 1 / 0 | 1 / 0 |
| GLS CAT II *(UAS IAP)* | 1 / 0 | 0 / 0 | 0 / 0 |
| **Total at all airports with UAS IAP** | **20 / 0** | **5 / 0** | **6 / 0** |
| **Total at P2 airports** | **20 / 4** | **31 / 5** | **28 / 8** |
| ILS CAT I | 17 / 4 | 31 / 5 | 28 / 8 |
| ILS CAT II | 2 / 0 | 1 / 0 | 3 / 0 |
| ILS CAT III *(UAS IAP)* | 9 / 0 | 1 / 0 | 1 / 0 |
| GLS CAT I *(UAS IAP)* | 1 / 0 | 0 / 0 | 0 / 0 |
| GLS CAT II *(UAS IAP)* | 0 / 0 | 0 / 0 | 0 / 0 |
| **Total at P2W airports** | **9 / 0** | **1 / 0** | **1 / 0** |

In Germany, a total of 41 airports have ILS/GLS approach procedures. An ILS CAT III approach is available at 20 airports. In addition to ILS CAT III, two German airports, Bremen (EDDW) and Frankfurt (EDDF), provide GLS CAT I procedures. Additionally, Frankfurt is the only German airport with GLS CAT II [35]. The only airports in California and Texas that have GLS procedures (CAT I at both) are Houston George Bush (KIAH) and San Francisco (KSFO).

Texas and California have about the same number of airports with ILS availability as Germany (see Table 4). The two US states have more P2 airports with ILS/GLS availability than Germany (36 in Texas and 36 in California versus 24 in Germany). However, Germany has more P2 airports providing UAS IAP (one in Texas and one in California versus nine in Germany).

### 4.2. UAS accessibility potential for regional aircraft at P2 airports

In the previous analysis on the potential of regional air cargo operations for UAS [20], regional aircraft with turboprop engines were the focus of the investigation. In the US, the Cessna 208 Caravan aircraft was the dominant cargo-only aircraft with more than 83% of domestic US cargo flight movements in 2021. In Europe, the ATR 42, ATR 72, and Embraer EMB 120 aircraft account for more than 94% of domestic European cargo flight movements by regional aircraft in 2021. Discussions with industry experts indicated that, in addition to regional turboprop aircraft, larger regional jet-powered aircraft may also be considered for UAS operations. Previous research by the German Aerospace Center (DLR) investigated the development and validation of a concept for the operation of unmanned cargo as part of the "Unmanned Freight Operations" (UFO) project between 2014 to 2017 [42]. In that work, different aircraft were analysed covering three use cases: express freight (Boeing 777F), company internal transport (Cessna 208), and disaster relief flights (no specific aircraft type). However, as discussed in Section 2.2, current efforts focus on using fixed-wing aircraft in the RAM realm at relatively small and under-utilized airports that typically do not service widebody aircraft such as a Boeing 777F. Hence this study was limited to regional aircraft, as defined in Section 2.2.

---

[9] Unless otherwise specified, data for the United States includes Puerto Rico and other US territories. A flight from Miami, Florida to San Juan, Puerto Rico, for example, would be counted as domestic.

### 4.2.1. Types of regional aircraft eligible for UAS

It was assumed that domestic flights have the highest potential for initial UAS operations because different countries are likely to have different regulations regarding UAS operations. Table 5 provides an overview of aircraft types used for domestic flight movements at P2 airports [16, 17].

**TAB 5** Domestic flight movements by regional aircraft in Europe and the US in 2022

| Domestic flight movements by regional aircraft | | Europe[a] | US |
|---|---|---|---|
| All operations[b] | Total flights | 246,796 | 1,313,204 |
| | ATR 42 | 6.9% | 0.7% |
| | ATR 72 | 16.9% | 0.6% |
| | Bombardier CL-600 (jet) | 15.0% | 0.2% |
| | Bombardier Dash 8-100 | 52.6% | 1.0% |
| | Embraer EMB 120 | 1.4% | -[d] |
| | Embraer ERJ 145 (jet) | 1.8% | 19.5% |
| | Cessna 208/208B | - | 23.5% |
| | Cessna 402 (piston) | - | 5.8% |
| | Beech 18[c] | - | 0.9% |
| | Canadair RJ200 (jet) | - | 22.9% |
| Cargo-only | Total flights | 10,529 | 155,266 |
| | ATR 42 | 23.6% | 1.6% |
| | ATR 72 | 43.4% | 3.3% |
| | Bombardier CL-600 (jet) | 8.7% | <0.1% |
| | Bombardier Dash 8-100 | <0.1% | - |
| | Embraer EMB 120 | 21.9% | -[d] |
| | Embraer ERJ 145 (jet) | - | - |
| | Cessna 208/208B | - | 56.0% |
| | Cessna 402 (piston) | - | 1.3% |
| | Beech 18[c] | - | 11.4% |
| | Canadair RJ200 (jet) | - | 0.6% |

a. Domestic in Europe refers to flight movements within each European country, summed over all European countries.

b. Refers to commercial flight movements with passengers and/or cargo on board.

c. FedEx Express has a waiver to report all of its small aircraft as Beechcraft Beech 18 C-185 (Beech 18) to the BTS, without regard to the actual aircraft type. Therefore, it will be excluded from further investigation throughout the study.

d. Ameriflight, a regional air cargo carrier, operates fourteen Embraer EMB 120 aircraft but is not included in the BTS database.

In Table 5, domestic cargo-only flight movements and flight movements with passengers and/or cargo on board (all operations) are compared. The regional aircraft in the table have turboprop engines, unless labelled (piston) or (jet). Note here that data are at the domestic level to give a more general picture of what type of regional aircraft are operating within different European countries versus the US. Significant differences in the total number of flights within European countries and the US are partially to not counting flights between European countries.

For domestic cargo-only flight movements in Europe, three turboprop aircraft types (ATR 42, ATR 72, and Embraer EMB 120) are again as dominant as in the previous 2021 analysis, with a combined total of just under 90% of the operations. In fact, the only jet aircraft type with a notable number of domestic cargo-only flight movements is the Bombardier CL-600 (Bombardier Challenger 600) aircraft that accounts for 8.7% of the operations in Europe (and 15% of all domestic operations in Europe). Cargo-only regional jet aircraft usage is even rarer in the US. Only 0.6% of cargo-only flights are operated by a single type of regional jet (Canadair RJ200). Conversely, the common aircraft in the US, the Cessna 208/208B and 402 or Beech 18 aircraft (see footnote d. in Table 5), are not used in Europe. Nonetheless, these regional aircraft types combined account for a significant share (68.7%) of cargo-only operations in the US.

Looking at the engine type of regional aircraft, Table 6 shows significant differences by the type of operation between regional jet aircraft and regional turboprop/piston aircraft (termed prop in Table 6) [16, 17].

**TAB 6** Flight movements by regional aircraft in 2022

| Flight movements by regional aircraft | | Germany | Texas | California |
|---|---|---|---|---|
| All operations[a] total[b] | prop | 18,521 | 17,112 | 21,911 |
| | jet | 45,565 | 111,108 | 60,459 |
| Cargo-only total[b] | prop | 9,225 | 11,544 | 13,546 |
| | jet | 55 | 229 | 8 |
| All operations[a] intra-state[c] | prop | 2,980 | 7,092 | 19,583 |
| | jet | 18,117 | 39,313 | 25,721 |
| Cargo-only intra-state[c] | prop | 112 | 7,058 | 13,540 |
| | jet | 2 | 25 | 0 |

a. Refers to commercial flight movements with passengers and/or cargo on board.

b. Refers to flight movements within the US and to intra- and extra-European flight movements.

c. Intra-state refers to flight movements within a US state and within Germany.

It is apparent that relatively few regional jet aircraft are used for cargo-only operations within Germany, Texas, or California, and that few are also used for cargo-only operations into and out of these areas. Rather, turboprop aircraft are predominant. However, jet aircraft are more common overall for all operations (commercial flight movements with passengers and/or cargo on board). Although flight movements with passengers on board are currently considered ineligible for conversion to UAS, the data show that intra-state flights, with regional flight distances of approximately <1,000 kilometers, with regional jets are common. Therefore, for this study, it was assumed that in the future, regional jet aircraft could be used for cargo-only UAS flights to serve under-utilized airports.

### 4.2.2. IAP availability at P2 airports

Table 4 shows that all P2W airports are towered across Germany, Texas, and California. Yet, non-towered airports are far more numerous than towered airports (see Section 3.1). To assess the availability of ILS/GLS (all CATs) and UAS IAP (only ILS CAT III and GLS CAT I and II), Table 7 breaks down the IAP by class of airspace and presence of air traffic control tower (towered) at P2 airports.

Table 7 shows that Germany has a significant number of regional airports in uncontrolled Class G airspace. However, of these 151 non-towered P2 airports, only four provide ILS procedures, and none have UAS IAP. There exist 22 towered P2 airports in controlled airspace, 20 of which have ILS or GLS (nine with UAS IAP).

**TAB 7** P2 airports by airspace class and IAP

| Airspace classes | Count of P2 airports | | | | | |
|---|---|---|---|---|---|---|
| | Germany | | Texas | | California | |
| All | 173 | | 376 | | 231 | |
| | 24 ILS /GLS | 9 UAS IAP | 36 ILS /GLS | 1 UAS IAP | 36 ILS /GLS | 1 UAS IAP |
| C | - | | 6 towered | | 3 towered | |
| | | | 6 ILS | 0 UAS IAP | 3 ILS | 1 UAS IAP |
| D | 22 towered | | 34 towered | | 41 towered | |
| | 20 ILS /GLS | 9 UAS IAP | 25 ILS | 1 UAS IAP | 25 ILS | 0 UAS IAP |
| E[a]/G | 151 non-towered | | 336 non-towered | | 187 non-towered | |
| | 4 ILS | 0 UAS IAP | 5 ILS | 0 UAS IAP | 8 ILS | 0 UAS IAP |

a. Germany does not operate airports in airspace Class E (see Table 1).

In the two US states analysed, Texas has 62.8% more P2 airports than California. Moreover, Texas has 117.3% more P2 airports than Germany. Looking at the share of non-towered airports, the results are again similar. Texas has 79.7% more P2 non-towered airports than California and 122.5% more than Germany. Both US states have only one P2W airport (Fort Worth Alliance, KAFW, in Texas and Fresno Yosemite International, KFAT, in California).

The visualization of all public airports, with P2 airports assigned a circle, including IAP configurations are shown in the following Figs. 1-3[10]. For each public airport, the highest possible IAP category is indicated with GLS being higher than ILS.

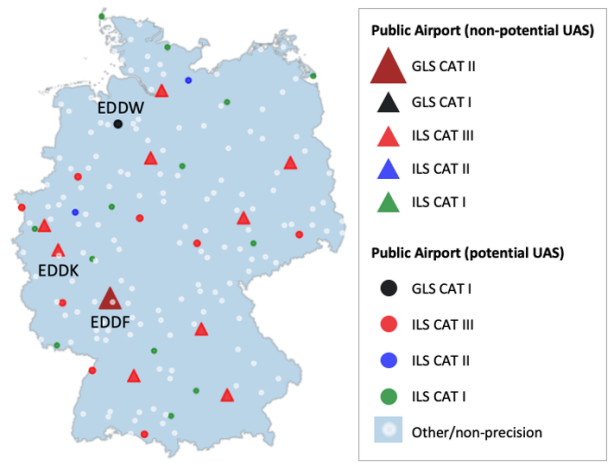

**FIG 1** Visualization of public airports in Germany with IAP availability

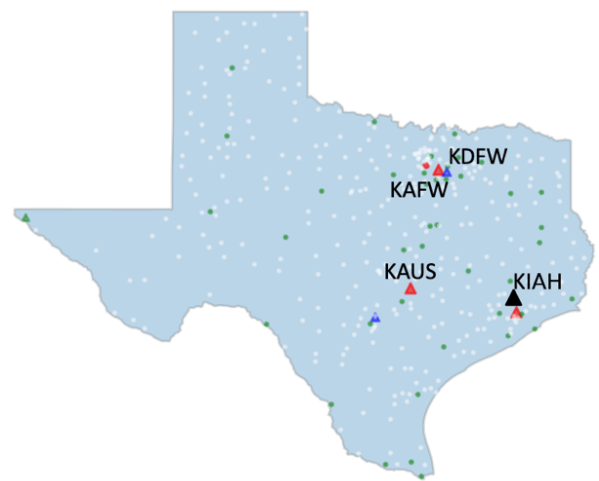

**FIG 2** Visualization of public airports in Texas with IAP availability

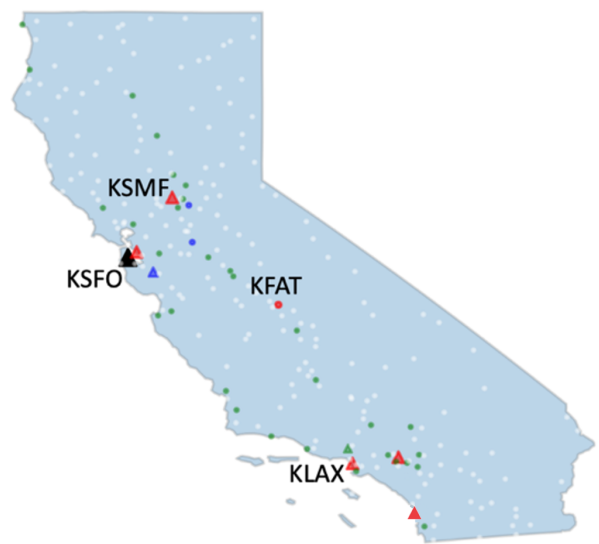

**FIG 3** Visualization of public airports in California with IAP availability

---
[10] Figs. 1-3 are not to scale with one another.

Figs. 1-3 show that many of the smaller airports are located closer to the areas with larger airports providing ILS CAT III and/or GLS close to the relatively larger cities. In Germany, there is a relatively high density of P2 airports in the west of Germany in the Rhine-Main region around Frankfurt (EDDF) and Cologne/Bonn (EDDK). In Texas, airport density around the metropolitan areas of Dallas-Fort Worth (KDFW), Austin (KAUS), and Houston (KIAH) is higher. California has a similar picture, where the density of smaller P2 airports increases around the metropolitan areas of Los Angeles (KLAX), San Francisco (KSFO), and Sacramento (KSMF).

### 4.2.3. Discussion of UAS accessibility potential for regional operations

After identifying regional aircraft types eligible for UAS operations and P2 airports in Germany, Texas, and California in the previous section, the next step is to analyse and discuss the accessibility potential of these regional aircraft at these P2 airports. For this analysis, regional aircraft are classified based on their operational empty weight (OEW[11]) and MTOW in tonnes (t). As regional aircraft have a wide variety of payload tonnage, the range between OEW and MTOW was considered for the UAS accessibility assessment to give a feasible range. According to a regional cargo industry expert, regional aircraft are often volumetrically filled before the aircraft's MTOW is exceeded. Therefore, if the OEW and MTOW of an aircraft is less than or equal to the rated gross weight capacity of the airport runway for the aircraft's wheel configuration, it was included in the accessibility assessment of the respective airport. UAS accessibility of regional aircraft is differentiated between total number of P2 airports as well as between towered (twrd) and non-towered (ntwrd) P2 airports.

Table 8 provides an overview of the most widely used regional aircraft types in Europe and the US (see Table 5) that are likely to be eligible for UAS operations and their accessibility potential at P2 airports. The OEW and MTOW in tonnes of each aircraft are listed in the column "Aircraft types" after the regional aircraft types. The metrics were used for the following regional aircraft type variants: ATR 42-600 (ATR 42) [43], ATR 72-600F (ATR 72) [44], Bombardier Challenger 650 (CL-600) [45], Bombardier DHC-8 Q200(-100) (Dash 8-100) [46], Embraer EMB 120 Brasilia (EMB 120) [47], Embraer ERJ 145 EP (ERJ 145) [48], Cessna 208 Caravan with cargo pod (C 208) [49], Cessna 208 Grand Caravan with cargo pod (C 208B) [50], and Canadair RJ200 ER (CRJ200) [51].

Taking the ATR 72 with an OEW of 11.80 tonnes and an MTOW of 23.00 tonnes as an example, this regional aircraft type can serve a total of 36 to 61 German[12] P2 airports, depending on how much usable fuel and payload is carried. Based on the rated gross weight capacity of the runways, 61 P2 airports allow an aircraft with MTOW of >10.50 tonnes (with the next higher airport MTOW being 12.00 tonnes) and 36 P2 airports allow an aircraft with MTOW of >20.00 tonnes (with the next higher airport MTOW being 25.00 tonnes) at which the ATR 72 would be allowed to operate in Germany. For each regional aircraft type analysed in Table 8, accessible German P2 airports (173 in total) include all 20 P2 towered airports with ILS/GLS, with nine of these P2 airports having a UAS IAP.

For the comparatively smaller regional aircraft types that are only used in the US for air cargo operations (e.g., Cessna 208), the Table 8 also indicates the number of German P2 airports that are eligible for fixed-wing UAS operations. However, it is not clear at present whether such aircraft would be utilized for cargo operations in Germany or Europe in the future.

**TAB 8** P2 airport accessibility by aircraft types eligible for UAS

| Aircraft types MTOW (OEW) | Count of accessible potential UAS airports MTOW (OEW) | | |
|---|---|---|---|
| | Germany | Texas | California |
| **ATR 42** 18.60 t (11.75 t) | **40 (61) total** 20 (21) twrd 20 (40) ntwrd | **66 (73) total** 36 (36) twrd 30 (37) ntwrd | **75 (104) total** 36 (37) twrd 39 (67) ntwrd |
| **ATR 72** 23.00 t (11.80 t) | **36 (61)** 20 (21) 16 (40) | **56 (73)** 35 (36) 21 (37) | **67 (100)** 35 (36) 32 (64) |
| **CL-600** 21.86 t (12.32 t) | **36 (59)** 20 (21) 16 (38) | **62 (73)** 36 (36) 37 (37) | **72 (100)** 36 (36) 36 (64) |
| **Dash 8-100** 16.47 t (10.48 t) | **40 (72)** 20 (22) 20 (50) | **72 (73)** 36 (36) 36 (37) | **75 (104)** 36 (37) 39 (67) |
| **EMB 120** 11.50 t (7.07 t) | **61 (76)** 21 (22) 40 (54) | **73 (74)** 36 (36) 37 (38) | **77 (118)** 36 (38) 41 (80) |
| **ERJ 145** 20.99 t (11.95 t) | **36 (61)** 20 (21) 16 (40) | **63 (73)** 36 (36) 27 (37) | **72 (100)** 36 (36) 36 (64) |
| **C 208** 3.63 t (2.21 t) | **148 (158)** 22 (22) 126 (136) | **267 (279)** 37 (38) 230 (241) | **192 (200)** 44 (44) 148 (156) |
| **C 208B** 4.00 t (2.41 t) | **146 (158)** 22 (22) 124 (136) | **266 (278)** 37 (38) 229 (240) | **192 (199)** 44 (44) 148 (155) |
| **CRJ200** 23.13 t (13.84 t) | **36 (59)** 20 (21) 16 (38) | **56 (72)** 35 (36) 21 (36) | **57 (67)** 33 (35) 24 (32) |

Overall, the analysis of current IFR flight movements in Section 3.1 shows that most of the flights are not operated at P2 airports today. The ten German towered airports that are not considered as P2 airports (Hub and IAA1) account for 87.7% of all annual IFR flight movements [27]. Similarly, a significant share of all IFR flight movements is operated at airports not considered as P2 airports in Texas (72.5%) and in California (78.7%) [29]. IFR flights are heavily concentrated at a few, large airports, supporting the assumption that there exist many under-utilized airports, many of which can be considered for initial UAS operations. Looking at the regional aircraft analysed, there are numerous different P2 airports in the investigated areas where an initial integration of fixed-wing UAS into the airspace system could be realized. Depending on the actual operating weight of the investigated regional aircraft based on its individual mission, a maximum of 158 P2 airports, mainly accessible by smaller turboprop aircraft (e.g., Cessna 208), and a minimum of 36 P2 airports would be accessible for fixed-wing UAS operations in Germany. In the US, a maximum of 279 and 200 P2 airports in Texas and California, respectively, would be accessible, again, mainly by smaller turboprop aircraft (e.g., Cessna 208/208B). On the other hand, a minimum of 56 and 57 P2 airports in Texas and California, respectively, would be accessible by regional aircraft. In this context, the share of P2 airports located in controlled and uncontrolled airspace varies. All three areas investigated have more P2 airports in uncontrolled airspace (non-towered airports) that are eligible for initial UAS operations.

---

[11] The OEW is the empty weight of an aircraft plus operational items including supplies necessary for full operations such as airline equipment and engine oil. Usable fuel that is needed to power the aircraft engines and the actual aircraft payload are excluded from the OEW.

[12] Some of the German airports impose operation hours and permits for MTOW operations. Upon request (PPR: Prior Permission Required), airports can be opened for air transport services outside of normal operating hours and for MTOW operations.

## 5. ANALYSIS OF INDIVIDUAL HIGH P2 AIRPORTS

This section investigates and compares individual airports in Germany, Texas, and California that have the highest potential to be utilized as P2 airports for the initial introduction of cargo UAS operations based on their runway MTOW allowances and current air transport operations. Here, both P2W and P2N airports can be considered as high P2. High P2N airports might need to be retrofitted with UAS IAP or other landing technologies (thereby making that airport a high P2W airport) first to enable widespread cargo UAS operations.

### 5.1. Current operations at (non-)P2 airports

As introduced in Section 3.2, P2W airports are likely to have a higher potential to be utilized for initial UAS operations than P2N airports. Nine P2W airports provide UAS IAP in Germany; Texas and California have one such airport each (Table 4). Given the relatively low number of airports that have the potential to be used for initial UAS operations with current certified landing systems in Germany and the two US states, many P2N airports will need a retrofit of UAS IAP or other landing technologies in the future to enable widespread fixed-wing UAS operations. P2N airports could be upgraded with certified landing technologies such as ILS CAT III or GLS as well as with landing technologies that are currently under development such as vision-based landing systems. In addition to P2W airports, P2N airports with appropriate runway MTOW allowances and commercial air cargo operations (that could be replaced by cargo UAS, for example) are defined as high P2 airports having a high potential for the introduction of initial UAS operations (i.e., high P2N airports) in the following sections.

Table 9 gives an overview about the commercial air transport operations at the main airports (non-P2 airports because they have annual IFR flight movements percentages >2.2%), the P2W airports, the P2N airports, and all other airports. The commercial air transport operations at these airports are distinguished by enplaned cargo in tonnes and enplaned passengers handled during annual flight movements, as well as by all-operations flight movements (all ops flight mov) in 2022 [17, 38, 39].

In Germany, over 90% of enplaned cargo and passengers are handled at the ten main airports. Accordingly, in Germany, between 4 and 5% are handled at the nine P2W airports. A similar picture is seen in Texas and California, where over 78% of enplaned cargo is operated at seven main airports in Texas and over 92% is handled at eleven main airports in California.

**TAB 9** Total commercial air transport operations at (non-)P2 airports in 2022

| Air transport operations at airports | Enplaned cargo (t)[a, b] | Enplaned passengers | All ops flight mov[c] |
|---|---|---|---|
| **Germany** | | | |
| Total main airports[d] | 4,919,953 (95.6%)[e] | 152,114,000 (91.8%) | 1,374,303 (47.6%) |
| Total P2W airports | 223,220 (4.3%) | 7,471,780 (4.5%) | 124,195 (4.3%) |
| Total P2N airports[f] | 1,150 (<0.1%) | 248,579 (0.2%) | 224,208 (7.8%) |
| Total other airports[g] | 310 (<0.1%) | 5.939.636 (3.6%) | 1,163,902 (40.3%) |
| **Total combined** | **5,144,633** | **165,773,995** | **2,886,608** |
| **Texas** | | | |
| Total main airports[d] | 1,844,497 (78.1%)[e] | 174,839,029 (96.2%) | 1,580,645 (91.2%) |
| Total P2W airports | 377,719 (16.0%) | 7,845 (<0.1%) | 22,911 (1.3%) |
| Total P2N airports[f] | 134,310 (5.7%) | 6,708,421 (3.7%) | 122,655 (7.0%) |
| Total other airports[g] | 4,222 (0.2%) | 283,297 (0.2%) | 8,269 (0.5%) |
| **Total combined** | **2,360,748** | **181,838,592** | **1,732,480** |
| **California** | | | |
| Total main airports[d] | 4,862,023 (92.7%)[e] | 190,776,761 (95.5%) | 1,673,377 (91.3%) |
| Total P2W airports | 14,438 (0.3%) | 2,155,276 (1.1%) | 25,125 (1.4%) |
| Total P2N airports[f] | 338,285 (6.4%) | 6,853,761 (3.4%) | 124,395 (6.8%) |
| Total other airports[g] | 30,077 (0.6%) | 70,971 (<0.1%) | 10,731 (0.6%) |
| **Total combined** | **5,244,823** | **199,856,769** | **1,833,628** |

a. Enplaned cargo on board cargo-only, belly freight, or combi freight flights.
b. Cargo consists of both freight and mail.
c. Flight movements refer to the sum of an arrival and departure for all national and international commercial flights that are both scheduled and non-scheduled.
d. Main airports refer to airports with annual IFR flight movements percentages ≥2.2% for the given area (country/state).
e. Percentage of total combined airport operations for the given area (country/state).
f. The listing of total P2N airports only includes airports that had >0 tonnes of enplaned cargo in 2022.
g. Other airports include P2N airports that did not have commercial air cargo operations in 2022 and all other airports such as private and military use airports. This data does not exclusively contain commercial flight data from fixed-wing aircraft but also from aerial vehicles such as from helicopters and piloted balloons. This affects especially all operations flight movements at the "total other airports" category.

Whereas the absolute number of enplaned passengers and all operations flight movements at the main airports is relatively comparable among the three investigated areas, the absolute number of enplaned cargo in tonnes varies among the areas. Germany (4.92 million tonnes) and California (4.86 million tonnes) have a similar amount of enplaned cargo operated at the main airports, more than double that of Texas (1.85 million tonnes). Note that the US numbers may be undercounted because Ameriflight, a major regional air cargo carrier based in Texas, is not included in the BTS data.

With respect to the enplaned cargo at P2W airports, Texas clearly dominates (377,719 tonnes,at KAFW alone). This high number is due to the fact that KAFW is a significant cargo hub for both FedEx and Amazon. Germany's total enplaned cargo at the nine P2W airports (223,220 tonnes) is almost entirely at Frankfurt-Hahn (220,127 tonnes) and California's lone P2W airport, KFAT[13] (14,438 tonnes) has a much lower volume of cargo.

Looking at the P2N airports that handled commercial air cargo, Germany has a comparatively low absolute number (1,150 tonnes) and share (>0.1%) of enplaned cargo compared to Texas (134,310 tonnes with 5.7%) and California (338,285 tonnes with 6.4%).

### 5.1.1. Individual high P2W airports

To identify high P2W airports to assess the potential of future UAS operations for air cargo missions, airports are ranked by enplaned cargo in tonnes that are handled at these airports. The more enplaned cargo currently handled at an airport, the more likely it can be assumed that the initial introduction of cargo UAS will start at those airports.

Table 10 lists all P2W airports in Germany, Texas, and California ranked by their enplaned cargo in tonnes. Data are sorted by operations by aircraft of all different sizes (all aircraft) and by aircraft that meet the definition of regional aircraft (<25 tonnes MTOW) [17, 38]. Enplaned cargo can be considered one of the main indicators of whether cargo UAS are eligible candidates for replacement of current operations. However, if an airport already has a comparatively high amount of flight movements by all operations but a low amount of enplaned cargo, there might be potential for expansion of air transport operations handled by cargo UAS at that airport in the future (i.e., increased cargo service to that airport).

Airports in Texas and California that are in controlled airspace providing traffic separation service by ATC are marked as towered (twrd) airports followed by their airspace class in Table 10. German airports in controlled airspace are marked as CTR (as introduced in Section 2.1, a CTR is controlled Class D airspace).

All airports listed in Table 10 are found to be suitable for regional cargo UAS operations in terms of regional aircraft accessibility, as certified landing technologies are already in place, airports are in controlled airspace, and the airports have a MTOW allowance that exceeds the MTOW of the regional aircraft investigated in this study. Based on current air cargo operations, Table 10 shows that Frankfurt-Hahn (EDFH) in Germany and Fort Worth Alliance (KAFW) in Texas stand out with the highest amount of annual enplaned cargo among the investigated areas. However, only 0.05% of enplaned cargo are transported by regional aircraft at EDFH and 4.11% at KAFW. These small percentages are partially explained simply by the fact that a large cargo jet (e.g., a Boeing 767) can carry significantly more tonnage than a regional cargo aircraft. Nonetheless, a comparison of the flight movements by all commercial air transport aircraft to those by regional aircraft shows that EDFH only has only ~2 flight movements by aircraft <25 t MTOW per day. By comparison, KAFW has ~15 such flights per day.

**TAB 10** P2W airports ranked by enplaned cargo in Germany, Texas, and California in 2022

| Commercial air transport operations at airports | Enplaned cargo (t) | | All ops flight mov | |
|---|---|---|---|---|
| P2W airport | All aircraft | Aircraft <25 t MTOW | All aircraft | Aircraft <25 t MTOW |
| **Germany** | | | | |
| Frankfurt-Hahn (EDFH) (CTR-D) | 220,127 | 114 | 13,264 | 668 |
| Karlsruhe/Baden-Baden (EDSB) (CTR-D) | 1,784 | 1,768 | 21,089 | 12,742 |
| Erfurt-Weimar (EDDE) (CTR-D) | 933 | 12 | 2,873 | 1,664 |
| Bremen (EDDW) (CTR-D) | 290 | 90 | 18,656 | 5,129 |
| Dresden (EDDC) (CTR-D) | 61 | 1 | 11,425 | 2,324 |
| Muenster/ Osnabrueck (EDDG) (CTR-D) | 21 | - | 23,072 | 15,320 |
| Kassel-Calden (EDVK) (CTR-D) | 4 | - | 13,723 | - |
| Friedrichshafen (EDNY) (CTR-D) | - | - | 8,407 | 4,996 |
| Niederrhein (EDLV) (CTR-D) | - | - | 11,686 | 5,132 |
| **Total** | **223,220** | **1,985** | **124,195** | **47,995** |
| **Texas** | | | | |
| Fort Worth Alliance (KAFW) (twrd-D) | 377,719 | 15,520 | 22,911 | 5,477 |
| **California** | | | | |
| Fresno Yosemite (KFAT) (twrd-C) | 14,438 | 16 | 25,125 | 4,556 |

The German P2W airport with the second highest amount of enplaned cargo, Karlsruhe/Baden-Baden (EDSB), handles significantly less enplaned cargo than EDFH or KAFW, but over 99% are handled by regional aircraft at EDSB that are likely to be converted for the introduction of cargo UAS. Likewise, EDSB has the second-highest amount of all operations flight movements (cargo and/or passenger flight movements) after Muenster/Osnabrueck (EDDG), with over half of the flight movements operated by regional aircraft. It can be concluded that, although Germany has a handful of airports that could be used for the introduction of cargo UAS, most of the airports currently receive little to no enplaned cargo. Therefore, cargo handling infrastructure at these airports may need to be installed, though the investigation of specific cargo handling infrastructure and capabilities at specific airports is outside the scope of this work.

---
[13] Although KFAT hosts the California Air National Guard 144th Fighter Wing, among others, it is not considered as a joint-use airport in the official FAA database and therefore was included in our analysis.

In both Texas and California, only a single airport has the needed landing technology to enable cargo UAS operations.

Overall, all P2W airports can be considered relevant for the introduction of initial cargo UAS operations since the needed landing technologies at these airports are already available. In Germany, EDFH dominates the amount of enplaned cargo of all P2W airports. In the US, the only P2 airports in Texas and California both have a significant amount of enplaned cargo (377,719 tonnes at KAFW and 14,438 tonnes at KFAT) with over 22,000 annual flight movements and a significant share handled by regional aircraft (23.8% at KAFW and 18.1% at KFAT).

### 5.1.2. Individual high P2N airports

This section identifies P2N airports that have a high potential to be upgraded with UAS IAP or other needed landing technologies to enable initial cargo UAS operations. Airports in Germany, Texas, and California are ranked by their enplaned cargo in tonnes to identify airports with commercial air cargo operations for a potential cargo UAS replacement or expansion of operations.

Table 11 ranks all 17 P2N airports that had commercial air cargo operations in Germany in 2022 [38]. Airports located in uncontrolled airspace Class G that does not receive separation by ATC are marked as non-towered Class G airports (ntwrd-G or RMZ-G). Airports in uncontrolled airspace marked as RMZ-G and airports in controlled airspace marked as CTR-D allow for IFR approaches and can be considered to have a higher potential for the initial introduction of UAS since fixed-wing UAS are expected to operate under IFR [25]. As introduced in Section 2.1, an RMZ is specially created for IFR approaches at German airports in uncontrolled airspace Class G.

In 2022, 17 P2N airports handled commercial air cargo operations in Germany. Four of these airports are located on islands in the German North Sea, namely Juist (EDWJ), Wangerooge (EDWG), Borkum (EDWR), and Norderney (EDWY). However, these airports on the German islands have a MTOW allowance of just 5.7 tonnes. As indicated in Table 5, European regional aircraft with potential for cargo UAS applications (e.g., ATR 42 and 72, CL-600, EMB 120) start at an OEW of 7.07 tonnes with a MTOW of up to 23.00 tonnes (see Table 8). Accordingly, the four P2N airports located in the German North Sea are not considered to have a high initial potential for early regional cargo UAS use cases since the dominant regional cargo aircraft types eligible for UAS operations are not able to operate there.

Excluding the airports in the German North Sea due to their MTOW allowance, the remaining 13 P2N airports in Germany (569.1 tonnes of annual enplaned cargo) can be considered high P2N airports. Eleven of these airports can be assigned a higher potential for early cargo UAS operations based on their availability of a CTR or RMZ. Eight of these eleven airports have MTOW allowances of ≤20 tonnes that limit the maximum operating weight of the analysed regional aircraft in Table 8. However, based on this analysis, 13 airports can be identified as high P2N airports that have a comparatively high potential to be upgraded with UAS IAP or other needed landing technologies.

**TAB 11** P2N airports ranked by enplaned cargo in Germany in 2022

| Commercial air transport operations at airports[a] | MTOW allowance (t) | Enplaned cargo (t) | All ops flight mov[b] |
|---|---|---|---|
| Mannheim City (EDFM) (CTR-D) | 10.0 | 546.2 | 11,364 |
| Juist (EDWJ) (ntwrd-G) | 5.7 | 486.2 | 10,106 |
| Wangerooge (EDWG) (ntwrd-G) | 5.7 | 60.1 | 17,035 |
| Borkum (EDWR) (ntwrd-G) | 5.7 | 27.9 | 3,436 |
| Emden (EDWE) (RMZ-G) | 14.0 | 11.1 | 9,542 |
| Norderney (EDWY) (ntwrd-G) | 5.7 | 7.0 | 2,089 |
| Straubing (EDMS) (RMZ-G) | PCN 40[c] | 4.1 | 4,477 |
| Strausberg (EDAY) (RMZ-G) | 14.0 | 3.1 | 24,100 |
| Braunschweig-Wolfsburg (EDVE) (CTR-D) | PCN 52[c] | 1.5 | 11,805 |
| Moenchengladbach (EDLN) (CTR-D) | PCN 30[c] | 1.4 | 35,312 |
| Frankfurt-Egelsbach (EDFE) (ntwrd-G) | 20.0 | 0.5 | 40,459 |
| Hof-Plauen (EDQM) (CTR-D) | 14.0 | 0.5 | 2,456 |
| Siegerland (EDGS) (RMZ-G) | PCN 53[c] | 0.3 | 19,836 |
| Bautzen (EDAB) (RMZ-G) | PCN 44[c] | 0.2 | 7,514 |
| Schoenhagen (EDAZ) (RMZ-G) | 14.0 | 0.1 | 15,774 |
| Eisenach-Kindel (EDGE) (ntwrd-G) | 20.0 | 0.1 | 1,913 |
| Wilhelmshaven (EDWI) (RMZ-G) | 14.0 | 0.1 | 6,990 |
| Total | | 1,150.4 | 224,208 |

a. Data for operations by regional aircraft that have a MTOW <25 tonnes are not available.

b. All operations flight movements do not exclusively contain commercial flight data from fixed-wing aircraft but also from aerial vehicles such as from helicopters and piloted balloons.

c. The Pavement Classification Number (PCN) indicates the load-carrying capacity of the runway pavement of an airport.

The 22 high P2 airports in Germany are highlighted in Fig. 4 with nine being P2W airports and 13 being P2N airports. Since all P2W airports are towered, the P2N airports are distinguished by towered (twrd, denoted by a triangle) and non-towered (ntwrd, denoted by a circle) operations. Main airports (all towered) include all airports with annual IFR flight movements percentages >2.2% and are therefore considered non-P2 airports (see Section 3.1).

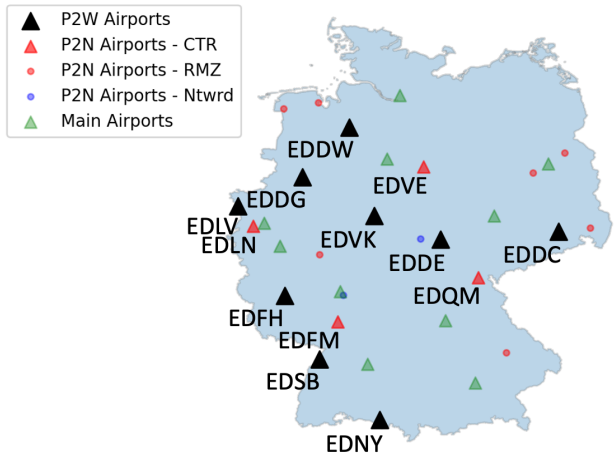

**FIG 4**   Visualization of P2 airports with and without UAS IAP (P2W and P2N, respectively), along with main, non-P2 airports in Germany

The German P2N airports can be distinguished by towered airports in controlled airspace Class D (i.e., airports having a CTR) and uncontrolled airspace Class G (i.e., airports having a RMZ or being non-towered). The eleven high P2N airports that either have a CTR or an RMZ, and therefore allow for IFR approaches likely to be required for initial UAS operations, are comparatively evenly distributed among Germany. However, the P2W airports having a CTR are more heavily located in the western part of Germany. In contrast, north-eastern Germany has few P2 airports.

Some of the high P2N airports (e.g., Schoenhagen Airport EDAZ and Strausberg Airport EDAY) are located near relatively busy main airports (e.g., Berlin-Brandenburg Airport EDDB). Air cargo operations potentially performed by UAS at the latter could therefore fly to these smaller P2N airports, which would relieve the larger main airports.

Table 12 lists the top ten and the remaining twelve other P2N airports that had commercial air cargo operations in Texas in 2022 [17]. Each airport is indicated as towered (twrd) or non-towered (ntwrd) and the airspace in which it is located.

Texas has 22 P2N airports in operation that had commercial air cargo operations in 2022. In total, these airports in Texas operate significantly higher amounts of enplaned cargo than German P2N airports (134,310 tonnes versus 1,150 tonnes). Nineteen of these airports are located in Class C or D airspace. Due to the MTOW allowances at the airports (>5.66 tonnes) that exceed the MTOW of regional aircraft dominant in the US (e.g., C208/B with a MTOW of 3.63/4.00 tonnes), all P2N airports in Texas can be considered as having a high potential for initial UAS operations.

Figure 5 visualizes the 23 high P2 airports with and without UAS IAP in Texas (22 P2N airports plus one P2W airport). The top 10 P2N airports that are towered are almost all located in larger cities that are a several-hour drive from other cities. These airports may be good candidates for the introduction of UAS IAP to enable cargo UAS operations. Many of the other P2N airports with towers are located in either the Dallas-Fort Worth metroplex or along major highways in-between major cities. Another interesting note is that none of Houston (KIAH, KHOU), San Antonio (KSAT), Austin (KAUS), or El Paso (KELP) – four major metropolitan areas with main airports – have P2 airports. Some potential routes that could be serviced by cargo UAS are KAFW to the West Texas airports (KLBB, KMAF, KABI, and KSJT) or to airports in South Texas (KLRD, KMFE, KHRL, and KBRO).

**TAB 12**   P2N airports ranked by enplaned cargo in Texas in 2022

| Commercial air transport operations at airports | MTOW allowance (t) | Enplaned cargo (t) | All ops flight mov |
|---|---|---|---|
| Lubbock (KLBB) (twrd-C) | 77 | 51,750 | 18,354 |
| Laredo (KLRD) (twrd-D) | 86 | 32,950 | 10,030 |
| Valley (KHRL) (twrd-C) | 91 | 31,438 | 13,204 |
| McAllen (KMFE) (twrd-D) | 86 | 10,453 | 9,680 |
| Midland (KMAF) (twrd-C) | 91 | 3,833 | 18,952 |
| Del Rio (KDRT) (ntwrd-G) | 29 | 760 | 2,560 |
| Brownsville (KBRO) (twrd-D) | 77 | 703 | 4,940 |
| San Angelo (KSJT) (twrd-D) | 45 | 672 | 3,789 |
| Abilene (KABI) (twrd-C) | 73 | 669 | 5,155 |
| Brownwood (KBWD) (ntwrd-G) | 14 | 304 | 535 |
| Other P2N airports combined (12 airports) (>5.66 t) | | 780 | 35,251 |
| Total | | 134,310 | 122,450 |

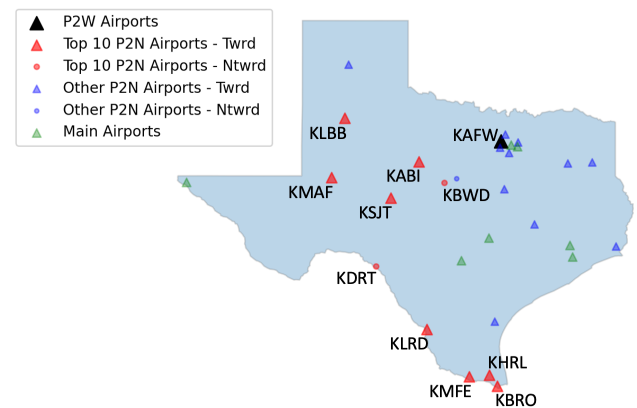

**FIG 5**   Visualization of P2 airports with and without UAS IAP (P2W and P2N, respectively), along with main, non-P2 airports in Texas

Table 13 lists the top ten and the remaining 30 other P2N airports that had commercial air cargo operations in California in 2022, distinguishing between towered (twrd) and non-towered (ntwrd) airports and their related airspace [17].

**TAB 13** P2N airports ranked by enplaned cargo in California in 2022

| Commercial air transport operations at airports | MTOW allowance (t) | Enplaned cargo (t) | All ops flight mov |
|---|---|---|---|
| San Bernardino (KSBD) (twrd-D) | 120 | 212,306 | 8,784 |
| Sacramento (KMHR) (twrd-D) | 127 | 68,156 | 3,064 |
| Stockton (KSCK) (twrd-D) | 68 | 48,175 | 2,823 |
| Santa Barbara (KSBA) (twrd-C) | 73 | 2,039 | 17,436 |
| Visalia (KVIS) (ntwrd-E) | 45 | 1,097 | 1,667 |
| Santa Maria (KSMX) (twrd-D) | 82 | 1,030 | 1,821 |
| Imperial (KIPL) (ntwrd-E) | 36 | 757 | 4,072 |
| Redding (KRDD) (twrd-D) | 58 | 742 | 5,044 |
| Bakersfield (KBFL) (twrd-D) | 70 | 731 | 6,086 |
| San Luis Obispo (KSBP) (twrd-D) | 67 | 685 | 11,103 |
| Other P2N airports combined (30 airports) (>5.44 t) | | 2,567 | 62,495 |
| Total | | 338,285 | 124,395 |

California has 40 P2N airports with commercial air cargo services in 2022. In total, these airports have higher volumes of enplaned cargo (338,285 tonnes) handled than airports in Texas (134,310 tonnes) and Germany (1,150 tonnes). Like Texas, all these airports in California can be considered high P2N airports due to the MTOW allowances at all of the airports (>5.44 tonnes) exceeding the MTOW of regional aircraft dominant in the US. Eighteen of these airports are located in controlled airspace Class C and D. The 41 high P2 airports with and without UAS IAP (40 P2N airports plus one P2W airport) in California are visualized in Fig. 6.

Like Texas, many of the top 10 P2N airports that are towered in California are in cities hours away by truck from major metropolitan areas. California overall has more P2 airports. Like Texas, few are near the main airports. California is a very mountainous state, and many of the major metropolitan areas along the western coast are hemmed in by mountains, leaving only a few overland routes to the smaller communities away from these areas. As such, route distances that might be driven by truck in a flatter location (e.g., much of Texas) are flown due to the mountainous terrain. This terrain, along with California's large population, has led to a robust regional air cargo network. However, this same terrain may cause difficulties with reliable cargo UAS command and control links, hindering introduction. The most likely initial area for introduction of cargo UAS could be the Central Valley, a large agricultural region in the center of the state. Possible routes here could be KFAT to surrounding communities.

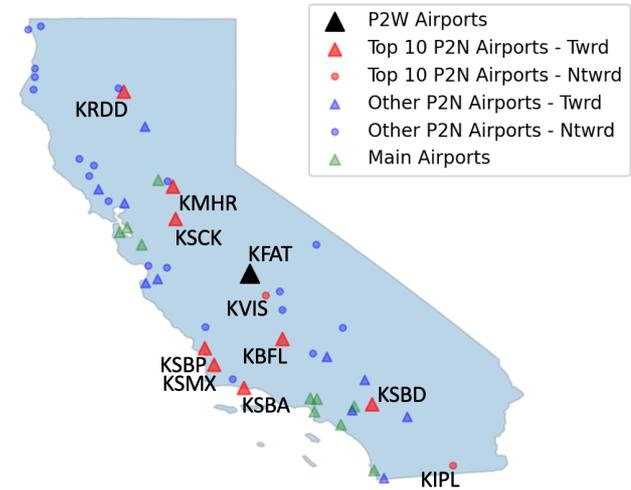

**FIG 6** Visualization of P2 airports with and without UAS IAP (P2W and P2N, respectively), along with main, non-P2 airports in California

### 5.2. Discussion of high P2 airports suitable for initial cargo UAS operations

Germany, Texas, and California each present unique challenges and opportunities for the introduction of regional air cargo UAS. In terms of cargo tonnage at those airports most able to accept UAS (i.e., P2W airports), Texas (377,719 tonnes) has significantly more tonnage than Germany and California combined (223,220 and 14,438 tonnes, respectively). However, Texas currently has only one P2W airport, meaning that at least one additional airport would need to have the appropriate technology for flights between airports to occur. Similarly, California also has only one P2W airport. Both states do, however, have a healthy demand for cargo across several P2N airports, with Texas' 22 such airports receiving 134,310 tonnes of cargo in 2022 and California's 40 such airports receiving 338,285 tonnes. With the introduction of needed IAP/landing systems, existing air cargo traffic in these states could be converted to UAS. One can conclude that, in Texas and California, it is the IAP/landing systems that are lacking, whereas the cargo handling infrastructure is likely at many of the airports already.

Conversely, Germany has nine P2W airports and 17 P2N airports. Two of the nine German P2W airports can be highlighted for the introduction of cargo UAS based on the total amount of enplaned cargo and enplaned cargo by regional aircraft. Frankfurt-Hahn (EDFH) handles 98.6% of all enplaned cargo of the nine German P2W airports and Karlsruhe/Baden-Baden (EDSB) handles 89.1% of all enplaned cargo operated by regional aircraft. None of the other airports received more than 1,000 tonnes of cargo in 2022 (and EDSB only barely passed that threshold). In fact, the 17 P2N airports combined received two orders of magnitude less cargo (1,150 tonnes) than similar airports in Texas. Thus, in Germany, there is much less of an existing regional air cargo route network. The introduction of cargo UAS in Germany is made easier due to the greater number of P2W airports but is hampered by a lack of existing regional air cargo and, possibly, the accompanying cargo infrastructure at airports.

Across all three areas investigated, all P2W airports have a high potential for initial cargo UAS operations because currently certified landing technologies likely for initial fixed-wing UAS operations are already available. On the one hand, a comparatively high amount of current enplaned cargo at P2W airports, such as at Frankfurt-Hahn (EDFH) in Germany, Fort Worth Alliance (KAFW) in Texas, and Fresno Yosemite International (KFAT) in California, could indicate the potential of these airports for the initial introduction of cargo UAS via one-to-one replacement of operations. On the other hand, if a P2W airport has a comparatively low amount of enplaned cargo, a high amount of all operations flight movements, such as at the German airports EDSB, EDDW, and EDDG, it could indicate the relevance of these airports due to their high volume of air transport operations with potential for expansion of services via cargo UAS.

The 13 high P2N airports in Germany have significantly less enplaned cargo handled (569 tonnes) than the 22 high P2N airports in Texas (134,310 tonnes) and the 40 high P2N airports in California (338,285 tonnes). Thus, the relevance of the 13 German high P2N airports for an upgrade with UAS IAP or other landing technologies appears quite small compared to the amount of enplaned cargo that is operated at the high P2N airports in Texas and California. Nevertheless, 11 of the 13 German high P2N airports are located in controlled airspace or have an RMZ that allows for IFR approaches likely to be required for UAS operations.

However, highly automated cargo UAS operations, especially for regional use cases with the availability of many under-utilized airports, could become relevant in Germany in the future, as air transportation is used for high-value and short-time-frame deliveries. This makes air transport a critical part of the freight infrastructure, despite its low tonnage percentage [52]. Even though the entire air cargo transport was only 0.1% of total freight tonnage transported in Germany in 2021 [53], Germany is an important country for intra- and extra-European logistics due to its central geographical location in Europe and excellent ground and air infrastructure. In Germany, freight transport is currently dominated by road and rail transport, which accounted for a combined 87.1% of total freight tonnage transported in 2021. Although air cargo traffic is relatively small compared to freight transport and passenger traffic by road and rail, it is important for overall economic performance [52]. Its importance could increase as freight traffic in Germany is expected to grow by 40% by 2030 compared to 2010 [54] and highly automated aircraft operations, such as cargo UAS, might create viable business cases.

## 6. CONCLUSION AND FUTURE WORK

Regional aircraft eligible for UAS operations and their accessibility potential at airports were analysed using 2022 data to assess the integration potential of regional fixed-wing cargo UAS into the airspace system. This study builds on previous research that identified Germany, Texas, and California as suitable areas for an initial integration of regional cargo UAS due to their relatively high number of smaller airports and/or current air cargo traffic. This paper investigates operations of regional piston, turboprop, and jet aircraft to identify airports suitable to serve regional aircraft eligible for UAS. All airports in Germany, Texas, and California were analysed according to their current IAP, with those procedures best suited to initial fixed-wing UAS operations (i.e., ILS CAT III or GLS), termed UAS IAP, given special attention. Emphasis was also given to the investigation of less busy airports (i.e., P2 airports), as it is anticipated that cargo UAS will initially start operating from under-utilized airports.

To establish a baseline for the comparative analysis of different areas, airports were defined as P2 airports if they provide public air transport services and have <2.2% IFR flight movements of all towered airports in the country/state. Additionally, all non-towered airports were classified as P2 airports. The total number of P2 airports with public air transport services was identified, with 173 in Germany, 376 in Texas, and 231 in California. However, currently, only nine P2 airports in Germany, one in Texas, and one in California provide UAS IAP availability. In the future, it is likely that P2 airports without UAS IAP will be equipped with GLS rather than ILS CAT III for UAS operations, since only one GLS installation per airport is required, as opposed to one installation per runway end, like ILS CAT III. This analysis shows that there is currently a dearth of P2 airports equipped with UAS IAP. Either more UAS IAP will need to be installed, or other landing technologies, such as vision-based technologies, will need to be developed to enable UAS accessibility at many under-utilized airports. Should other landing technologies be developed, however, the results of this study indicate that future fixed-wing UAS could access a high number of P2 airports, regardless of powerplant.

Based on runway MTOW allowances, current air transport operations at airports, and airspace classes, individual high P2 airports were identified in Germany, Texas, and California. Since only eleven airports in the investigated areas provide UAS IAP, individual high P2 airports are distinguished by availability of UAS IAP. High P2 airports without UAS IAP might be upgraded with UAS IAP or other landing technologies first to enable widespread cargo UAS operations. Among the investigated areas, Germany has 13 high P2 airports without UAS IAP, Texas has 22, and California has 40 that have a comparatively high potential for the retrofitting of ILS CAT III, GLS, or other needed landing technologies for fixed-wing UAS operations. Alternatively, should technologies onboard the aircraft advance such that, for example, a ILS CAT I or area navigation (RNAV) approach with vertical guidance could be used, this work showcases many high P2N airports at which cargo UAS operations could occur.

Although this study focused on UAS accessibility based on the availability of UAS IAP at airports, other challenges also limit UAS operations. Future work will attempt to quantify these limitations, including the availability of reliable command and control (C2) link performance, interactions with other IFR and VFR traffic, availability of contingency airports, and plans to mitigate the loss of the C2 link. The analysis presented in this paper will also provide inputs to fast-time simulation studies, whereby different percentages of current regional air cargo operations may be replaced with UAS operations and extended to additional routes operated by UAS.